# Spin Rotation Invariant Spin Triplet Superconducting liquids

Fei Zhou

*ITP, Minnaert building, Leuvenlaan 4, 3584 CE Utrecht, The Netherlands*

(November 2, 2018)

Spin ordering and its effect on the low energy quasiparticles in a p-wave superconducting fluid are investigated. We study the properties of a new $2D$ quantum spin triplet superconducting liquid where the ground state is spin rotation invariant. In quantum spin disordered cases, zero energy skyrmions are topologically stable because of a fermionic zero mode in space-time monopoles which serve as a quantum protectorate. The low energy quasiparticles are bound states of the bare Bogolubov- De Gennes ($BdeG$) quasiparticles and skyrmions, which are charge neutral bosons at the zero energy limit. Further more, the spin collective excitations are fractionalized ones carrying a half spin and obeying fermionic statistics. In thermally spin disordered superconducting states, a bare $BdeG$ has an infinity energy and each two $BdeG$ quasiparticles form spin zero bound states.

## I. INTRODUCTION

Quantum number fractionalization has been a fascinating subject. The curiosity about this subject has been widely shared by mathematical physicists, field theorists and condensed matter physicists [1–10]. The earliest example of the quantum number fractionalization is perhaps the angular momentum fractionalization of an electron when confined with a Dirac monopole [1]. As now known to many of us, the electron carries a half integer angular momentum in this case, by contrast to what we've learned in atomic physics.

Zero modes in certain topological excitations are vital to quantum number fractionalization. This was first appreciated in a 1D field theory model where it was observed that fermionic excitations in a charge conjugate model can carry one half of the elementary quantum number [2]. A powerful method developed later on provides further insights into the nature of the problem [5]. The authors of [5] showed that in a continuum theory, the quantum number of a topological excitation is a continuous function of the parameters in the theory and can be an arbitrary fraction in the absence of a charge conjugation symmetry [5]. These ideas were also deeply rooted in the beautiful theories on the index of the Dirac operators which were put forward both before and after the discoveries in field theories [12]. The mechanism leading to the fractionalization in these cases is the spectral flow in the presence of a singular gauge transformation. The spectral flow, especially the level crossing in the field theory furthermore results in remarkable ideas of Witten's global anomaly, the Skyrmeon model of nucleaon etc [13].

The statistics of these exotic objects in spatial dimensions higher than one was investigated at the same period of time. A spinless bosonic particle bound with a monopole has half spin and behaves as a fermion and vice verus [4,3,14]. In this situation, statistics transmutation takes place which further complicates the issue. The authors of [15] first introduced $\theta$- statistics in two-dimension in their *tour de force* analysis of two-body wavefunctions with arbitrary boundary conditions. Later on, the possibility of having anyons in $2D$ was discovered, which can be achieved by a flux attachment and which now becomes a very powerful icon [16]. In an interesting paper, Wilczek and Zee further emphasised that the spin and statistics of a spin texture can be related to the linking number of two trajectories in the target space; the linking number is identical to a Hopf invariant of a rotating skyrmion and results in anyonic skyrmion excitations [17].

The interest on the quantum number fractionalization in condensed matter systems was very much prompted by the remarkable discoveries of charge one but spinless quasiparticles and spin one but chargeless quasiparticles in one dimensional polyacetylene [6]. The experiment discoveries of fractional quantum Hall effects in 2DEG and the theory on FQH incompressible liquids with quasiparticles carrying one third of an electron charge [7,8] further stimulated investigations on this subject. This phenomenon inpired many theoretical works on fractional statistics. And the later theoretical investigations on the mechanism of high temperature superconductivities, where the important concept of spin-charge separation was introduced [9–11], were once more focusd on this issue. In the backdrop of FQHEs in 2DEG and high temperature superconductivities in cuprates, two intensively studied subjects in condensed matter physics in past two decades, many efforts have been made to achieve a better understanding on the issue of quantum number fractionalization [18–28].

The theory on one dimension polyacetylene was established more than two decades ago [6] and is worth a detailed discussion here because the quasiparticles in 2d p-wave superconductors here reminisce, in terms of spirits, the mid-gap states discovered in that paper. There are at least two essential ingredients in those one dimensional systems which make the fractionalized excitations possible. And we want to emphasis both of them.

In one dimensional polyacetylene of $CH$-polymer, each carbon atom has to share with the neighboring atoms three $\sigma$ electrons and one $\pi$ electron. The $\sigma$ electrons form covalent bonds and only $\pi$ electrons are conducting and responsible for the electronic properties of the polyacetylene. So the conduction band is half filled which is subjected to a Peierls instability. A lattice distortion takes place; carbon atoms at the $2n + 1$-th sites along the chain are displaced by a $0.2A$ in one direction while atoms at the $2n$-th sites are displaced towards an opposite direction. A charge density wave with a period of two lattice constants, $2a$ is formed. Simultaneously, a Peierls gap opens up right at the fermi surface leaving behind an incompressible electron liquid where the valence band is completely filled and separated from the empty conduction band by the Peierls gap.

The ground state obviously has twofold degeneracy when the Peierls instability takes place. In a dimerized picture, two degenerate configurations A and B can be transformed into each other by a displacement of a lattice constant. This suggests kink-like solitons in the polyacetylene, which are the domain-walls separating A and B configurations. A careful calculation shows that the size of these kinks can be rather small, of a few lattice constants implying a small mass for the topological excitations. The analysis carried out by Su, Heeger and Schrieffer convincingly demonstrates that when a kink like soliton is inserted into the chain, the fermionic spectrum which is otherwise fully gapped, starts to develop a mid-gap state. Electrons of each spin specie in the valence band contribute a half mid-gap state so that there is precisely single electronic state with two fold spin degeneracy appearing at the fermion surface. The existence of topological excitations, particularly that of the zero modes hosted in these excitations is the first ingredient we wish to emphasis.

When the mid-gap states are doubly occupied or empty, the corresponding quasiparticles are spinless but carry a unit negative or positive charge. When the mid-gap states are singly occupied, the quasiparticles carry no charge but with spin one half. The charge and spin conservation are guaranteed by the conservation of topological winding numbers. For a give finite system, one always has to insert a kink-anti kink pair. When the kink carries spin one half (up) but zero charge, the antikink has to carry spin down and zero charge. However, the kink can also carry a unit positive charge but zero spin while the antikink carry a unit negative charge and zero spin.

It is also obvious that quasiparticles are weakly interacting and well-defined only when the topological objects, in this case, the kinks and anti-kinks don't have long range interactions. This is in fact a necessary condition for fractionalization to take place and is the second ingredient we are turning to. In a mean field approximation which turns out to be a remarkably good approximation in this case, the two fold degeneracy inherited in the Peierls instability ensures the absence of a long range interaction and a liberation of kink-anti kink pairs. However, this is not so when the degeneracy is absent. For instance in polythiophene, a cousin of polyacetylene, the two-fold degeneracy is lifted. The kink and anti-kink interact with a potential which grows linearly as a function of distance between them. This is a sign of confinement of kinks as well as the fractionalized quasiparticles discussed above. In a most general situation, the interaction between fractionalized objects can be highly nontrivial. Especially in many magnetic systems, including the one we are going to address in this paper, fractionalization takes place only when a special mechanism of quantum protection exists. And this quantum protectorate functions only under certain topological enviornments, mostly through conserving certain topological charges or ordering of topological fields.

To demonstrate the idea of fractionalization in magnetic systems, we turn to a brief discussion of 1D antiferromagnetic spin chain, which has been a subject of extensive investigations and is relatively better understood than other higher dimension systems. For this purpose, we address this issue in the context of the AKLT formulation of the problem so that a close analogy can be drew between the spin chain and one dimension charge conjugate polymer discussed above [29]. For spin one half chain with Majumdar-Ghosh Hamiltonian

$$H = \sum_i P_{3/2}(\mathbf{S}_i + \mathbf{S}_{i+1} + \mathbf{S}_{i+2}) \qquad (1)$$

where $P_{3/2}$ is the projection operator for a spin-3/2 state, the ground state can be viewed as a collection of spin singlet pairs formed between adjacent sites. If each site emits one monomer standing for a one-half spin, the ground state can be represented by a periodical array of dimers connecting each pair of two adjacent sites, or a valence-bond crystal so to speak, with a period of two lattice constants. The ground state has twofold degeneracy, the configurations of which (A and B) can be obtained by displacing one array of dimers by a lattice constant. The excitations are fully gapped and the spin system is incompressible.

A spin one half excitation can be created by breaking a dimer into two monomers and moving one of the monomers to the infinity while keeping the other one fixed in the bulk of the spin chain. So again, a monomer separating the A and B configurations corresponds to a kink; this time, one half spin is hosted in the kink. What one can easily show is that kinks are liberated in this case and spin-1/2 are free excitations because of the two fold degeneracy in the problem. This is not surprising at all given that we started with spin-1/2 chain.

Let us now turn to a less trivial example examined by AKLT. Consider a spin-1 chain with the following Hamiltonian

$$H = 2\sum_i P_2(\mathbf{S}_i + \mathbf{S}_{i+1}) - (\beta + \frac{1}{3})(\mathbf{S}_i \cdot \mathbf{S}_{i+1})^2. \qquad (2)$$

At $\beta = -1/3$, the ground state again is a valence bond crystal; two monomers are emitted from each site and terminated at two different neighboring sites. Each two adjacent sites are connected by one dimer forming a VBC with a period of one lattice constant. The ground state is nondegenerate and the excitations are fully gapped.

Creation of spin 1/2 excitations, which carry only half of the spin of the underlying particles at each site, involves breaking a dimer and sending one of the monomer to infinity as emphasised above. But, because the ground state is nondegenerate, this procedure unavoidly provoks some higher energy configurations, say a string of dimerized pairs(see Fig.1). In a dimerized pair configuration, two monomers emitted from each site always end at a same site so that a singlet is formed between two adjacent sites. A string of dimerized pairs emitted from a spin 1/2 excitation, or a monomer can be terminated either at the infinity or at another spin 1/2 excitation. This suggests a confinement of spin-1/2 excitations in spin chains with underlying spin equal to one and we are not able to create a free excitation which carries half of the underlying spin. This is again not surprising.

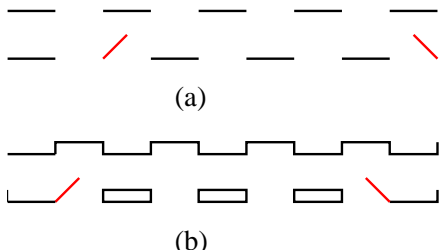

Fig.1 Creation of spin-1/2 excitations in 1D VBC states in a spin one half chain a) and a spin one chain b). In b), to create two spin 1/2 excitations (monomers), a dimerized string (down figure) has to be introduced to the ground state (up figure).

What is important and interesting is that the energy difference of a dimerized pair configuration and the VBC can be adjusted by varying $\beta$. As $\beta$ approaches a large positive value, the energies of dimerized configurations are lowered and the dimerized states are stablized. A large negative value of $\beta$ on the other hand stablizes the VBC state. It is therefore expected that these two configurations have to be degenerated at a certain point, mostly likely at $\beta = 1$. When this happens, the confining potential disappears encouraging a spin fractionalized excitation with half spin according our argument. This oversimplied argument, which doesn't take into account fluctuations etc nevertheless demonstrates the key idea behind fractionalization. And a surprise doesn't come at no cost. A spin-1/2 excitation in a spin chain of underlying spin equal to 1 can be observed only at certain points ($\beta = 1$ for example) out of the whole parameter space. This time we are also lucky enough to get the right answer as far as the spin of the excitations is concerned regardless the crude approximation in our reasoning. More sophisticated calculation using Bethe ansatz solutions does confirm the existence of these objects [30]. This is further supported by $k = 2$ conformal field theory calculations [31].

The picture just demonstrated in this short discussion plays a central role in the phenomena of fractionalization. It is evident that the fractionalized spin excitations are "hosted" in some form of solitons, this time, kink like ones. The topological aspect, especially the attachment of a high energy "string" to some naively fractionalized excitations, indicates a possible catastrophic consequence of having such an excitation. The issue of fractionalization therefore is intimately associated with the termination of such a long range singularity, by dynamical means. This was also the idea behind the early work of Kivelson, Sentha and Rokhsar, Rokhsar and Kivelson on the 2D antiferromagnetic spin-1/2 systems in square lattices, and most recently Moessner and Sondhi on 2D frustrated antiferromagnetic spins in triangle lattices [18,21,27].

In the quantum dimer model in square lattices, again each dimer represents a singlet pair. To make contact with spin liquids of half spins, one also has to apply a hard core constraint so that only one link emanating from a site is occupied by a dimer. The Hamiltonian is

$$H = -J(|=><|||+h.c.) + V(|||><|||+|=><=|) \quad (3)$$

where the first term characterizes the energy gain from the resonance between two parallel configurations of dimers (horizontal and vertical pairs). The last term stabilizes each pair of the parallel dimers when $V$ is negative but distablizes such a parallel configuration when $V$ is a positive. The competition between them results in two distinct phases.

When $J \ll V$, the respulsion between parallel dimers is dominating and a VBC phase which contains no parallel dimers emerges as a result of the repulsion. When $J \gg V$, a column phase where dimers sit on the top of each other prevails. In between, it is believed that at $J = V$, an equal amplitude resonating valence bond liquid exists as a point where, crudely speaking, a column configuration and a VBC are degenerate. Each phase has fourfold ground state degeneracy as one can easily visualize and both are fully gapped and spin incompressible.

In both phases, the creation of a monomer, or a spin-1/2 excitation provoks a disturbance of the crystal long range order, even with four degeneracy. The disturbance which is caused by a frustration in any dimension higher than 1D, is long range by nature. It again can be represented by a string which is emitted from a monomer or half spin in the bulk and ends only at the boundary or another half spin. We are coping precisely with the same situation as in 1D spin 1 chain. We therefore intend to conclude that spin-1/2 excitations, sometimes coded as spinons in the literature are not generic objects in this model and only exists at one point $J=V$ when the ground state is an equal amplitude superposition of a number of configurations. Otherwise, only spin one excitations similar to gapped spin wave excitations can be observed though at each lattice site the elementary spin is one half. This surprising result that the excitations carry spins doubling the spins of underlying particles, causes major frustrations over the early believe that spinons with half spins are elementary excitations in the system.

However, as recently pointed out, this frustration can be partially removed when the 2D antiferromagnetic systems are geometrically frustrated, say due to geometries such as triangle lattices. In that case, the authors of [27] were able to demonstrate that in a fraction of the parameter space the long range order found in either VBC or a Column phase is absent in ground states. This relieves the frustration of creating a monomer in the bulk of the system and makes spinons possible. We will not pursue further here but refer to the original work.

Existence of spinions in other 2D antiferromagnetic systems was also suggested in a few later works [32,33]. In [32], the authors showed the free spinon excitations in a crossed-chain model of a 2D spin liquid; in [33], the authors investigated the stability of a fractionalized phase in a kagome lattice and arrived at conclusions consistent with those in [27].

The other powerful way to describe the fractionalization in magnetic systems is to introduce gauge fields to characterize various induced interactions. And this is also the language we are going to employ in the discussion of quasiparticles in this paper. We will give a pedagogical summary of the development and refer to original works when the details are concerned. The power of the field theory in the study of quantum magnetism perhaps was first fully appreciated in a work on one dimensional spin chain [34]. In this pioneering work, a topological term was derived to distinguish the spin chain of integer and half integer spins, especially the excitation spectra. The Haldane gap observed in 1D integer spin chains but not in half integer spin chain is a manifest trophy of the field theory approach. However, a massive employment of the gauge field based approach in the study of many-spin systems perhaps was triggered by a series of papers of Anderson et al. [9,10] on the mechanism of high temperature superconductivities, see for example [19,35–39]. It was envisioned that an antiferromagnetic insulator under doping first becomes a resonating valence bond state(RVB), a collection of spin singlet pairs, which eventually is responsible for the superconductivity; further more, spinons likely form a fermi surface even in the charge insulating limit. The RVB theory indicates neutral spin-1/2 fermionic spinons and so the existence of these spinon excitations which is a key feature in the now widely accepted mechanism becomes the most crucial issue [9,10,18].

However, an analysis of the spin disordered states using the field theoretical approach reveals a rather frustrating aspect of the problem. That is, when the long range Neel order disappears as the superconductivity emerges, the spin fluctuations effectively lead to $U(1)$ gauge fields. And spin-1/2 but bosonic spinons always carry unit charges with respect to the induced $U(1)$ fields, reflecting the Berry's phase a spinon develops under a spin rotation. So they interact with each other via $U(1)$ gauge fields. The key feature of the induced $U(1)$ field is that in $2D$ even at zero temperature, they always mediate a confining force between two $U(1)$ charges. Namely, the energy of two spinons separated at distance $L$ is proportional to $L$ and this catastrophe manifests itself by forbidding any excitation carrying a $U(1)$ charge. Within the frame work of that approach, neither neutral fermionic spinons nor bosonic spin 1/2 spinons can exist. This is clearly reflected in the paper by Wiegmann [19]. An interesting phenomenological proposal to avoid this difficulty is to introduce a Hopf term to have deconfining gauge fields and statistical transmutations at the same time [40]. The other interesting but more recent proposal was made by Senthil and Fisher; for discussions on this $Z_2$ gauge field based approach, we refer to their original paper [26].

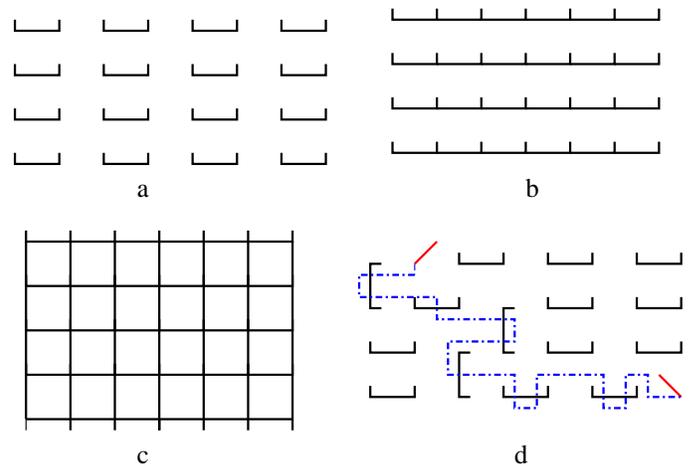

Fig.2 Following [20](see also [62]) in disordered limits, a) Fourfold degenerate VBC states for "spin" $S=1$ in a 2D square lattice; b) Twofold degenerate VBC states for $S=2$; c) Nondegenerate VBC states for $S=4$; d) The creation of spin-1/2 excitations in a 2D VBC state.

The Hopf-term-based proposal, promising and intriguing at the first sight was proved to be inappropriate in a antiferromagnetic system where the spatial parity is unbroken [62,63]. Attempts to derive the Hopf term micro-scopically in Heisenburg antiferromagnetic systems [62] indicate that instead there are other local Berry's phase terms which define the nature of disordered states. These Berry's phase terms result in destructive interferences between space-time monopole events at different points at square lattices. For odd integer spins all monopole events with unit charges are suppressed, and only monopoles with double charges are allowed; for half integer spins both unit charge and double charge monopole events are forbidden and monopoles are always quadrupled. The ground state in this disordered limit has fourfold degeneracy for half integer spins and twofold degeneracy for odd integer spins [62]. This fascinating result was later substantiated and refined in [20]. The authors of [20] introduced an $SU(N)$ representation and explored systematically the Berry's phase effect in the context of an effective compact $U(1)$ field theory. It was further pointed out that those degenerate states are of spin Peierls type breaking the lattice symmetry and correspond to the 2d valence bond crystal states closely connected with those early works of AKLT. They also emphasised that due to the confining property of the $U(1)$ gauge fields certain topological orders are absent and the bosonic spinon excitations are confined. In another interesting work, these authors examined a frustrated Kagome lattice and suggested that the spin one half spinons are deconfined due to a charge 2e Higgs field disregarding the underlying spins [25].

Though a Hopf term induced statistical transformation is believed not to take place in usual antiferromagnetic systems, because of the absence of time reversal symmetry breaking and parity breaking, theorists do agree and believe it should happen in systems such as chiral spin liquids [22]. Then one can expect that the spin 1/2 spinons can exist in the excitation spectrum. As we should see below, a $2D$ p-wave superconductor is precisely such an example. The existence of a topological term in p-wave superconducting states was previously demonstrated in [54]. Most attentions in those early works have been paid to many exotic properties of topological excitations in symmetry broken states. However, the issues of spin ordering and spin-phase separation haven't been explored; in particular, spin fluctuations' effects on quasiparticles remain to be understood, especially under the influence of the Hopf term. This will be the focus of this article.

In this paper, we will demonstrate the existence of spin rotation symmetry restored p-wave superconducting states and explore the properties of quasiparticles and collective excitations in these states. In section II, we scrutinize the spin ordering in the context of spin-charge separation, and prove the coexistence of a long range phase order and a short range spin order in a $2D$ spin triplet p-wave superconducting state. This spin rotation invariant state is characterized by a finite range spin correlation and $hc/4e$ vortices as the elementary topological excitations.

In section III, we characterize the topological orders in rotation invariant states. We show in a quantum spin disordered case the zero energy topological charges are conserved due to a quantum protectorate. In section IV, we examine the spin fluctuations' influences on excitations, particularly, quasiparticles and quantum collective spin excitations in $2D$ spin disordered p-wave superconductors. The elementary quasiparticles are Bogolubov-De Gennes ($BdeG$) quasiparticles hosted in gapless skyrmions and obey bosonic statistics in quantum disordered cases. The spin collective excitations are shown to be fractionalized ones carrying a half spin and obeying fermionic statistics, by contrast to the spin wave excitations in spin ordered p-wave superconducting states($SOpSS$s). In thermally disordered cases, $BdeG$ quasi-particles form bound states. Spin triplet superconducting states are believed to exist in $^3He$, many heavy-fermion superconductors and most recently layer perovskite $Sr_2RuO_4$ crystals [42–45]. We believe the interesting properties studied in this work can be observed in certain limits.

We should remark that the issue of fractionalization in spin triplet superconducting liquids was recently addressed by Demler et al. [41]. There are two important differences between the current work and that of [41]. First, we are going to cope with an order parameter living in space $[S^2 \times S^1]/Z_2$ while Demler et al. dealt with $[S^1 \times S^1]/Z_2$. Though from the point of view of induced $Z_2$ gauge fields which are the focuses of that work, this distinction is not important, it is however vital when spin textures and the Hopf terms are concerned. Second, the important aspect discussed in this paper, the Bogolubov quasi-particles hosted in skyrmions, depends crucially on the topology and is unique in the model studied here. The properties list in Tab.1. are novel ones of the spin disordered superconducting liquids. On the other hand, our theory is valid when the $Z_2$ gauge fields are frozen and $\pi$-disclinations are gapped. For a detailed discussion about $Z_2$ gauge fields and their effects on fractionalization, we should refer to the work by Demler et al. [41].

Fractionalization, though has been discovered in very diversified strongly correlated condensed matter systems which bear no similarities at first sight, appears to share one remarkable common feature. It is the topological excitations interacting with electrons one way or the other which make exotic quasiparticles or collective excitations possible. This also lies in the heart of earlier examples discovered in field theoretical models and mathematical physics [4,3,5,12]. Indeed, more sophisticated considerations have led us to believe that quantum number of exotic excitations seen in quantum Hall effects or appearing in theories on spin liquids should depends on the ground state degeneracy of the liquids in certain multiply connected manifolds. The connection between the ground state degeneracy and the fractional charges in quantum Hall liquids was raised in [24,22]; the topological orders in spin liquids were emphasised in a few occasions, see for instance [23,28]. As we will see, this notion plays a paramount role in our case as well.

## II. SPIN-PHASE SEPARATION AND SPIN CORRELATIONS

### A. Spin-phase separation

The term "spin-phase" separation is used in p-wave superconducting liquids to represent a separation of the phase dynamics and spin dynamics. So the low lying goldstone modes can be well classified as phasons and spin waves. However, this does not imply, at least not directly, the coexistence of short range spin correlation and a long range order in phases. In contrast, in the mean field approximation of BCS type, the "spin-phase" separation defined above always indicates a spin long range order if a phase rigidity is present. So the fact that we can have spin disordered superconducting liquids in 2D should not be taken for grant. And indeed, in any three dimension sample both the spin order-disorder transition and superconductor-metal transition take place at temperatures within the Ginsburg region and practically simultaneously.

To get oriented with these statements, let us first take a phenomenonlogical approach. We are going to consider the cooper pairs condensed in spin triplet states, or spin one states. The spin component of the pair wavefunction can be written in Wely-Majorana representations as

$$u(\mathbf{\Omega}_1)v(\mathbf{\Omega}_2). \quad (4)$$

Here $u(\mathbf{\Omega})$ and $v(\mathbf{\Omega})$ are expressed in terms of a vector $(u,v)$ on a unit sphere,

$$u(\mathbf{\Omega}) = \exp(i\chi)[\cos\frac{\theta}{2}\exp(-i\frac{\phi}{2})u_0 + \sin\frac{\theta}{2}\exp(i\frac{\phi}{2})v_0]$$
$$v(\mathbf{\Omega}) = \exp(-i\chi)[-\sin\frac{\theta}{2}\exp(-i\frac{\phi}{2})u_0 + \cos\frac{\theta}{2}\exp(i\frac{\phi}{2})v_0] \quad (5)$$

where $\mathbf{\Omega} = (\sin\theta\cos\phi, \sin\theta\sin\phi, \cos\theta)$. And the unit vector $(u,v)$ can be written as $u = \cos\theta_0/2\exp(i\phi_0/2)$, $v = \cos\theta_0/2\exp(-i\phi_0/2)$. The wavefunctions in Eq.4 form an overcomplete set fro spin-one states. The quantum spin nematic state which interests us below corresponds to a state $\mathbf{\Omega}_1 = \mathbf{\Omega}_2$. But to see how phase-spin separation takes place we keep two unit vectors independent at the moment.

In this representation, spin operator is defined as

$$\mathbf{S}^+ = u\frac{\partial}{\partial v}, \mathbf{S}^+ = v\frac{\partial}{\partial u},$$
$$\mathbf{S}_z = \frac{1}{2}[u\frac{\partial}{\partial u} - v\frac{\partial}{\partial v}] \quad (6)$$

which satisfies the usual commutation relationship $[\mathbf{S}_\alpha, \mathbf{S}_\beta] = i\hbar\epsilon_{\alpha\beta\gamma}\mathbf{S}_\gamma$. Spin $S$ states are expressed as degree $2s$ polynomials of $u$ and $v$; $u^{2S-m}v^m$, $m = 0, ...2S$. The scalar productor is defined as $\int f^*(u,v)g(u,v)d\Omega$.

As an exercise, one calculates the kinetic energy $E(\Omega, \chi) = (8\pi)^{-1}\int d\Omega_0 \; u^*(\mathbf{\Omega})\frac{-\nabla^2}{2m}u(\mathbf{\Omega})$ of the spin one half state $u(\mathbf{\Omega})$ as

$$E(\Omega, \chi) = \frac{1}{2m}|\nabla\chi - \mathbf{A}|^2 + \frac{1}{2m}|\nabla\Omega|^2.$$
$$\mathbf{A} = \cos\theta\nabla\frac{\phi}{2}, \mathbf{F}_{\mu\nu} = \frac{1}{2}\epsilon_{\alpha\beta\gamma}\mathbf{\Omega}_\alpha \cdot \frac{\partial\mathbf{\Omega}_\beta}{\partial x_\mu}\frac{\partial\mathbf{\Omega}_\gamma}{\partial x_\nu}. \quad (7)$$

where $\mathbf{F}_{\mu\nu} = \epsilon_{\mu\nu}\partial_\mu\mathbf{A}_\nu$.

Therefore, the charge degree of freedom $\chi$ is coupled to a topological gauge field $\mathbf{F}_{\mu\nu}$. It is this coupling that in general fails to yield a spin-charge separation in antiferromagnetic spin systems. Eq.7 also indicates that the current carried by a state $\mathbf{u}$ is

$$J(\mathbf{\Omega}) = \frac{1}{m}\big(\nabla\chi - \cos\theta\frac{\nabla\phi}{2}\big) \quad (8)$$

with the last term from a spin rotation. In terms of the current, a spin-phase coupling manifests in an extra contribution due to the spatially varying $\mathbf{n}$.

The supercurrent $J_s(\mathbf{\Omega}_1, \mathbf{\Omega}_2)$ carried by a paired state $\sqrt{6}u(\mathbf{\Omega}_1)v(\mathbf{\Omega}_2)$ can be obtained by calculating

$$J_s(\mathbf{\Omega}_1, \mathbf{\Omega}_2) = 6\int \frac{d\Omega_0}{4\pi}u*(\mathbf{\Omega}_1)v^*(\mathbf{\Omega}_2)\frac{i\nabla}{2m}u(\mathbf{\Omega}_1)v(\mathbf{\Omega}_2). \quad (9)$$

However, a direct calculation shows that $J = 0$ if $\Omega_1 = \Omega_2$ as in a quantum spin nematic state. Otherwise,

$$J_s(\mathbf{\Omega}_1, \mathbf{\Omega}_2) \approx \frac{1}{m}\nabla\chi + \frac{1}{m}\sin\theta[\delta\theta\nabla\phi - \delta\phi\nabla\theta] \quad (10)$$

with

$$\delta\theta = \frac{\theta_1 - \theta_2}{2}, \delta\phi = \frac{\phi_1 - \phi_2}{2},$$
$$\theta = \frac{\theta_1 + \theta_2}{2}, \phi = \frac{\phi_1 + \phi_2}{2}. \quad (11)$$

when $\delta\theta, \delta\phi \ll 1$. Or

$$J_s(\mathbf{n} + \frac{\mathbf{l}}{2}, \mathbf{n} - \frac{\mathbf{l}}{2}) = \frac{1}{2m}\epsilon_{\alpha\beta\gamma}\mathbf{n}_\alpha\mathbf{l}_\beta\nabla\mathbf{n}_\gamma. \quad (12)$$

From here one concludes that there is a coupling between the spin and charge sector. However, the current described in Eq.12 is much smaller than that in Eq.8 as $\mathbf{l}$ approaches zero. As a result, the effective gauge field in this case is

$$\mathbf{F}^p_{\mu\nu}(\mathbf{n},\mathbf{l}) = \frac{1}{2}\epsilon_{\alpha\beta\gamma}[\mathbf{n}_\alpha\frac{\partial\mathbf{l}_\beta}{\partial x_\nu}\frac{\partial\mathbf{n}_\gamma}{\partial x_\mu} + \frac{\partial\mathbf{n}_\alpha}{\partial x_\nu}\mathbf{l}_\beta\frac{\partial\mathbf{n}_\gamma}{\partial x_\mu}]. \quad (13)$$

It is negligible compared with the usual Pontryagin field in Eq.7 because of its dependence on $\mathbf{l}(\ll 1)$.

Indeed, a direct calculation of the kinetic energy of the cooper pairs condensed at a $u(\mathbf{\Omega} + \mathbf{l}/2)v(\mathbf{\Omega} - \mathbf{l}/2)$ state shows

$$E = \frac{\rho_s}{2m}|\nabla\mathbf{\Omega}|^2 + \frac{\rho_s}{2m}|\nabla\chi|^2 + \frac{\rho_s}{m}\nabla\chi \cdot \mathbf{A}^p(\mathbf{\Omega}, \mathbf{l}) \quad (14)$$

and $\mathbf{A}^p$ is the vector potential of $\mathbf{F}^p$ which vanishes as $\mathbf{l}$ approaches zero. $\rho_s$ is the superfluid density. In the lowest order gradient expansion, we conclude a spin-phase separation as far as the linear dynamics is concerned.

Having said this, we should also emphasis that the pairs which can be considered as spin one particles carry no $U(1)$ charges defined with respect to the Pontryagin field $\mathbf{F}_{\mu\nu}$, because of the *spin-phase* separation. As we will see later, on the other hand, the Bogolubov- De Gennes quasiparticles do carry unit $U(1)$ charges so that *spin-charge* separation does not take place.

Eq.14 further indicates that in the mean field theory, the phase rigidity always induces a spin rigidity and the superconductor- metal transition is always accompanied by an establishment of a magnetic order in $\mathbf{n}$. A more sophisticated approach should take into account the fluctuations in both spin and phase sectors. At a dimension higher than two, one can easily confirm that

$$\frac{T_c - T_c^{sm}}{T_c} \sim \frac{T_c - T_c^*}{T_c} \sim \frac{1}{\rho_0 \xi(0)^3} \quad (15)$$

where $T_c^{sm}$ is the superconductor-metal phase transition temperature, $T_c^*$ is the induced spin order-disorder transition temperature and $T_c$ is the mean field critical temperature. $\rho_0$ is the total density and $\xi(0)$ is the zero temperature coherence length. Eq.15 shows the subtlety of the spin-phase separation and consequently the coexistence of short range spin correlation with a long range phase order in a $3D$ system. The righthand side of the equation which virtually characterizes the size of the Ginsburge region is always much less than one. Therefore, in most of spin triplet superconducting states discovered so far, the mean field picture is more or less correct and from a practical point of view, the two transitions do happen at the same temperatures in a bulk system.

However, in general, the spin disorder and the phase disorder are driven by very different mechanisms and the two phase transitions belong to different universality classes. We should show that Eq.15 is false when a 2D p-wave superconducting liquid is concerned and a short range spin correlation does coexist with the superconductivity. For that purpose, we now turn to a microscopic derivation to include both the zero point motions of $\mathbf{n}, \chi$ and spectral flows in fluctuations.

### B. The effective action

We will be specifically interested in a p-wave superconductor with an order parameter

$$\hat{\Delta}(\mathbf{k}) = \Delta_0(T)(\sigma_2 \sigma \cdot \mathbf{n})(k_x + ilk_y)\exp(i\chi), \quad (16)$$

where $l = \pm 1$, and $\hat{\Delta}$ is defined in the space of spin $\uparrow, \downarrow$ [42,43]. $\mathbf{n}$ is a unit vector living in a two-sphere, $S^2$. This superconducting state breaks a $U(1)$ symmetry of the phase of the order parameter, a $U(1)$ symmetry of the rotation along $z$ axis and a $S^2$ symmetry of the rotation of spin quantization axises. In addition, it breaks the pairty invariance connected with $l = \pm 1$ and a time reversal symmetry.

One can always undo a rotation along the $z$ axis by a $U(1)$ gauge transformation and restoring this rotation symmetry also means an absence of superconductivities. We will not be interested in this symmetry restoring but focus on the rotation symmetry breaking or restoring in the spin sector within the superconducting phase. For this reason, we reserve the "rotation symmetry" solely for the $S^2$-symmetry in the spin sector. Consequently, the rotation invariant state suggested in the title refers to a state where the $S^2$ spin rotation symmetry is restored.

The corresponding Hamiltonian for the superconducting state in the Nambu space of $\Psi = (\psi^+, i\tau_2\psi)$ can be written as,

$$H = \sigma_3 \epsilon + \sum_{i=x,y} \sigma_i \{\partial_i, \hat{\Delta}\}_+, \quad (17)$$

where $\hat{\Delta}$ is defined as $\hat{\Delta} = \Delta_0 \exp(i\sigma_3\chi)(\mathbf{n} \cdot \tau)$ and $\epsilon(\mathbf{k}) = \hbar^2\mathbf{k}^2/2m - \epsilon_F$. We use $\sigma$ as the Nambu space Pauli matrix and $\tau$ as the spin space ones. We assume the spin-orbit interactions are weak and $\mathbf{n}$ is a unit vector in a sphere $S^2$. The internal space of the symmetry broken state is $\mathcal{R} = [S^1 \times S^2]/Z_2$, similar to that of Bose-Einstein condensates of $^{23}Na$ [48] studied recently. The order parameter observes a discrete symmetry: $\hat{\Delta}(\mathbf{n}, \chi) \to \hat{\Delta}(-\mathbf{n}, \chi + \pi)$ and represents a quantum spin nematic p-wave superconducting state.

To obtain an effective theory, we integrate over the fermionic degrees of freedom and make a gradient expansion. At low temperatures, we obtain the following results

$$\mathcal{L} = \mathcal{S}_{ab}\mathbf{P}_{sa}\mathbf{P}_{sb} - \mathcal{T}_0\Phi^2 + \frac{1}{8\pi}\left(\mathbf{E}^2 + \mathbf{B}^2\right)$$
$$+ \mathcal{S}_{ab}^{\alpha\beta}\nabla_a \mathbf{n}_\alpha \nabla_b \mathbf{n}_\beta - \mathcal{T}^{\alpha\beta}\partial_t \mathbf{n}_\alpha \partial_t \mathbf{n}_\beta + \frac{\mathcal{N}}{4\pi}\epsilon^{\mu\nu\lambda}\mathbf{W}_\mu^3 \mathbf{F}_{\nu\lambda}. \quad (18)$$

In Eq.18, $\mathbf{P}_s = \frac{1}{2}\nabla\chi + e\mathbf{A}^{em}$ and $\Phi = \frac{1}{2}\partial_t\chi + e\phi^{em}$ are the gauge invariant momentum and potential respectively. The superscript $em$ is introduced to distinguish the usual electric magnetic vector potential $\mathbf{A}^{em}$ from the topological field $\mathbf{A}$ defined in terms of $\mathbf{n}$ in Eq.7. And $\mathbf{W}_\mu^3$ is the vector potential of $\mathbf{F}_{\mu\nu}$ (defined in Eq.20).

$$\mathbf{F}_{\mu\nu} = \frac{1}{2}\mathbf{n} \cdot \partial_\mu \mathbf{n} \times \partial_\nu \mathbf{n}. \qquad (19)$$

In deriving Eqs.18,19, we have introduced two vector potentials defined in terms of a $U(1)$ rotation $U_c$ and a $SU(2)$ rotation $U_s$,

$$\mathbf{A}_{c\mu} = iU_c^{-1}u_c^{-1}\partial_\mu u_c U_c = \sigma_3(\mathbf{A}_\mu^{em} + \frac{1}{2}\partial_\mu\chi),$$
$$\mathbf{A}_{s\mu} = iU_s^{-1}u_s^{-1}\partial_\mu u_s U_s = \tau_\alpha \cdot \mathbf{W}_\mu^\alpha \qquad (20)$$

($\mu = 0, 1, 2$, stands for coordinates in $1 + 2$ dimension space.). $u_c = \exp(-i\sigma_3\frac{\alpha(t)}{2})$, and $u_s = \exp(i\tau_3\frac{\beta(t)}{2})$.

The time dependent gauge rotations $\alpha(t), \beta(t)$ are introduced to preserve the anti-periodic boundary conditions for Fermions under a nontrivial gauge transformation [49]. And the two gauge rotations are defined as

$$U_s^+ U_s = 1, U_s^+ \mathbf{d} \cdot \tau U_s = \tau_3,$$
$$U_c^+ U_c = 1, U_c^+ \sigma_i \exp(i\sigma_3\chi)U_c = \sigma_i. \qquad (21)$$

At last, the temperature dependences of all coefficients are given in Appendix B. Let us define $\mathcal{S}_{ab}^{\alpha\beta} = \delta^{\alpha\beta}\mathcal{S}_{ab}$, and dimensionaless coefficients

$$\mathcal{T}_0 = \nu_0\tilde{\mathcal{T}}_0, \mathcal{S}_{ab} = \frac{\rho_0}{2m}\tilde{\mathcal{S}}_{ab}, \mathcal{T}_{ab} = \delta^{\alpha\beta}\nu_0\tilde{\mathcal{T}}. \qquad (22)$$

$\nu_0$ is the averaged density of states at the fermi surface of the normal state, $\rho_0$ is introduced as the density; $m$ and $\epsilon_F$ are the mass and the fermi energy respectively. Then $\tilde{\mathcal{T}}_0, \tilde{\mathcal{S}}_{ab}, \tilde{\mathcal{T}}$ and $\mathcal{N}$ are calculated as

$$\tilde{\mathcal{T}}_0 = \frac{1}{4\nu_0}\int \frac{d^2k}{(2\pi)^2}F_2(\mathbf{k})\tan(\frac{E(\mathbf{k})}{2kT}),$$
$$\tilde{\mathcal{S}}_{ab} = \frac{2m}{\rho_0}\int \frac{d^2k}{(2\pi)^2}\mathbf{v}_a\mathbf{v}_b F_1(\mathbf{k}),$$
$$\tilde{\mathcal{T}} = \frac{1}{\nu_0}\int \frac{d^2k}{(2\pi)^2}F_1(\mathbf{k});$$
$$\mathcal{N} = \int \frac{d^2k}{4\pi}\epsilon^{\alpha\beta\gamma}\frac{\mathbf{M}_\alpha}{E^3(\mathbf{k})}\frac{\mathbf{M}_\beta}{\partial k_x}\frac{\mathbf{M}_\gamma}{\partial k_y}F_1(\mathbf{k}). \qquad (23)$$

$E(\mathbf{k}) = \sqrt{\epsilon^2(\mathbf{k}) + \mathbf{k}^2\Delta_0^2/k_F^2}$, and $\Delta_0$ is the temperature dependent gap. The director $\mathbf{M}$ is defined as $\mathbf{M}(\mathbf{k}) = (v_\Delta \mathbf{k}_x, v_\Delta \mathbf{k}_y, \epsilon(k))$; $v_\Delta = \Delta_0/k_F$. And $F_{1,2}$ are calculated in appendix B.

Following Eq.23, $\mathcal{T}_0\nu_0^{-1}$ is a unity at $T = 0$ and varies smoothly as a function of the temperature [48], $\nu_0$ is the averaged density of states at the fermi surface. $\mathcal{S}_{ab}$ on the other hand approaches zero at the mean field critical temperature $T_c$. When the director $\mathbf{n}$ is a planar vector confined in the $\theta = \pi/2$ equator, i.e. $\mathbf{n} = (\cos\phi, \sin\phi, 0)$, the spatial gradient terms coincide with previous results for the $A$ phase of $^3He$ [50,51]. The action is valid when the frequency and the wave vector are smaller than $\Delta_0(T)$ and $\xi_0^{-1}(T) = \Delta_0(T)/v_F$ respectively.

Eq.18 suggests a few important properties of the quantum spin nematic p-wave superconductors. First of all, the dynamics of spin $\mathbf{n}$ and phase $\chi$ is completely separated at the low frequency limit(except the entanglement due to the $Z_2$ group in the functional integral [41,48,52,53], which we will not discuss in this paper.). It reflects a spin-phase separation in the p-wave superconductors, as argued in the previous section on a phenomenonlogical ground. Second, for an isotropic fermi surface which interests us in this article, $\mathcal{S}_{ab} = \delta_{ab}\rho_s(T)/2m$, and $\rho_s(T)$ is the temperature dependent superfluid density which vanishes at the critical temperature $T_c$. So the spin and phase dynamics are characterized by an $O(3)$ $\sigma$-model (NL$\sigma$M) and an $xy$ model respectively. At the mean field approximation, $\mathbf{n} = \mathbf{e}_z$ and $\chi = $ const.. This corresponds to a conventional $SOpSS$. There are three Goldstone modes; two of them are spin waves $= (1, \pm i, 0)$ with a linear dispersion. In an isotropic case, $\tilde{\mathcal{S}}_{ab} = \delta_{ab}\tilde{\mathcal{S}}$; the spinwave velocity is $v_s(T) = v_F\sqrt{\tilde{\mathcal{S}}/2\pi\tilde{\mathcal{T}}}$. And the last mode is the usual plasma wave, with a dispersion $\omega = \sqrt{2\pi e^2\rho_0 k/m}$ in $2D$ at $T = 0$.

The topological term was previously derived in [54,55]. There, $\mathcal{N}$ is effectively the number of skyrmions living in the $(\mathbf{k}_x, \mathbf{k}_y)$ plane if we define a skyrmion configuration in the plane as $\mathbf{M}(\mathbf{k}) = (v_\Delta\mathbf{k}_x, v_\Delta\mathbf{k}_y, \epsilon(k))$. Namely, $\mathcal{N} = \oint_{S_F}\frac{d^2k}{4\pi}E^{-3}(\mathbf{k})\epsilon^{\alpha\beta\gamma}\mathbf{M}_\alpha\partial_{k_x}\mathbf{M}_\beta\partial_{k_y}\mathbf{M}_\gamma$ and $\mathcal{N} = 1$ for a single band case. This term determines the topological orders in the fields $F_{\mu\nu}$ and defines the structure of quasiparticles. We investigate the spin ordering, quasiparticles and topological excitations based on Eq.18.

The topological term arises naturally in a system where the parity and time reversal symmetry are broken. It is therefore not suprising similar terms can be present in the effective theories of other unconventional superconductors where these symmetries are broken. Some implications of this phenomenon on (spin) Hall effects were indeed explored recently in a series of papers [56–59]. We refer to these original papers for a detailed discussion on transports.

### III. SPIN DISORDERED SUPERCONDUCTING LIQUIDS

Because of an extra branch of Goldstone modes in the spin sector, the spin order is more fragile than the phase order in the problem when fluctuations are taken into account. This provides a possibility of having spin disordered p-wave superconducting states ($SDpSS$s) where the $S^2$ symmetry is restored and only the $U(1)$ symmetry is broken. Such a state which is spin rotation invariant in nature differs from the conventional $SOpSS$s where the $S^2$ symmetry is broken and there is a long range order in $\mathbf{n}$.

The finite temperature phase diagrams of the $O(3)$ Nonlinear $\sigma$ Model (NL$\sigma$M) were previously analyzed in great details [60]. In the current situation, just as the superfluid velocity $\rho_s(T)$, all coefficients in the action, $\mathcal{S}_{ab}^{\alpha\beta}, \mathcal{T}_{ab}, \mathcal{S}_{ab}, \mathcal{T}$ depend on temperatures because of quasiparticle excitations. Taking this into account, we arrive at the following results in $2D$.

**QSDpSS** When $\Delta_0 \gg \epsilon_F$, the spin order is spoiled by quantum fluctuations and there is a short range spin correlation even at zero temperature with a finite correlation length. [1] This state will be named as a quantum $SDpSS$, or $QSDpSS$.

**TSDpSS** When $\Delta_0 \ll \epsilon_F$, the spin order is established at zero temperature and the correlation length

$$\xi_2 = \frac{v_s(T)}{\Delta_s(T)}, \Delta_s = T \exp\big(-\frac{2\pi[\rho_s(T)/2m - \Gamma]}{T}\big) \quad (24)$$

is finite only at finite temperatures (in a saddle point approximation). Here $\Gamma \sim \Delta_0(T)$. This state is a thermally $SDpSS$, or $TSDpSS$. Discussions about quantum orders, zero energy skyrmions and $BdeG$ quasi-particles will be carried out paralelly for both cases.

### A. Instantons and their suppressors

The nature of spin fluctuations in a $QSDpSS$ can be explored in a spinor representation of Eq.18. By introducing $\eta^+ \tau \eta = \mathbf{n}$, $\eta = (\eta_1, \eta_2)^T$ and $\eta^+ \eta = 1$, we obtain for $\eta$ the following Lagrangian in $SDpSS$s,

$$\mathcal{L}_\eta = \frac{1}{2f^2}|(i\partial_\mu - \mathbf{A}_\mu)\eta|^2 + \frac{\Delta_s(T)}{\Delta_0(T)}\eta^+\eta + \frac{\mathcal{N}}{4\pi}\epsilon^{\mu\nu\lambda}\mathbf{A}_\mu\mathbf{F}_{\nu\lambda}. \quad (25)$$

And $\eta$ is a bosonic field carrying a unit charge with respect to $\mathbf{A}$ fields and spin $1/2$. In Eq.25, $2f^2 = 2m\Delta_0/\sqrt{\tilde{\mathcal{S}}\tilde{\mathcal{T}}}\rho_0$; we have introduced the following rescaling: $\tau \rightarrow it\xi_0/v_s$, $\mathbf{r} \rightarrow \mathbf{r}\xi_0$ and Eq.25 is written in a $(2+1)$ Euclidean space.

When the spins are disordered, the gapped spin fluctuations induce an effective Maxwell term. Upon integrating over spin wave excitations, based on a general consideration of the gauge invariance and the parity symmetry breaking, we conclude the NL$\sigma$M should be reduced to

$$\mathcal{L}_s(\mathbf{F}_{\mu\nu}) = \frac{1}{2g}\mathbf{F}_{\mu\nu}\mathbf{F}_{\mu\nu} + \frac{\mathcal{N}}{4\pi}\mathbf{W}_\mu^3\mathbf{F}_{\mu\nu}... \quad (26)$$

$g(\Delta_s)$ is a function of $\Delta_s$ the spin gap. A direct diagrammatic calculation yields the following estimate for $g$

$$\frac{1}{g} = \frac{\Delta_0}{8\pi\Delta_s}. \quad (27)$$

In $(2+1)$ space $\mathbf{x} = (\tau, \mathbf{r})$, it is convenient to introduce a field, $\mathbf{H}_\eta = \frac{1}{2}\epsilon^{\eta\mu\nu}\mathbf{F}_{\mu\nu}$. $\mathbf{H}_\tau = \mathbf{F}_{xy}$ represents $U(1)$ magnetic fields along $z$ direction, $\mathbf{H}_x = \mathbf{F}_{y\tau}$ and $\mathbf{H}_y = \mathbf{F}_{\tau x}$ are the $y, x$-components of the electric fields. $\mathbf{H}$ is the solution of the Gauss equation

$$\partial_\eta \mathbf{H}_\eta(\mathbf{x}) = \sum_m Q_m \delta(\mathbf{x} - \mathbf{x}^m) \quad (28)$$

in the presence of space-time monopoles $\{\mathbf{x}^m\}$ in $(2+1)d$ Euclidean space.

Following Eqs.26, 28 one can obtain an effective action in terms of monopoles' coordinates. However, the Hopf term leads to destructive interferences between rotated skyrmions terminated at space-time monopoles. All monopole events are therefore completely suppressed. This is reflected in the action for monopole-like instantons. But before presenting the result, let us characterize the topological structures of these instantons.

Consider a rotating skyrmion terminated at the origin in a $S^2 \times S^1$ space

$$\mathbf{n}(\rho, \phi; \tau) = \big(\sin\theta\cos\tilde{\phi}, \sin\theta\sin\tilde{\phi}, \cos\theta\big) \quad (29)$$

where

$$\tilde{\phi} = Q\phi - \gamma(\tau),$$
$$\theta(\rho, \tau) = 2arcos\frac{\rho}{\sqrt{\rho^2 + \tau^2}}\Theta(\tau),$$
$$\gamma(\tau + L_\tau) - \gamma(L_\tau) = N2\pi. \quad (30)$$

To obtain a desired boundary condition for $\eta$-quanta and quasi-particles discussed in the next section, we also require $\gamma(\tau + L_\tau) - \gamma(\tau) = N2\pi$. The perimeter along $\tau$ direction $L_\tau$ is finite for the discussion of a finite temperature case; $L_\tau = T^{-1}$.

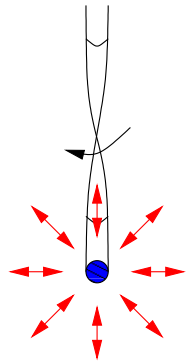

---

[1] This is supported by a mean-filed calculation in a $CP^N$ representation.

Fig.3 Skyrmions terminated at space-time monopoles at different points carry different phases $\gamma_B$, which result in destructive interferences.

The gauge fields in this "texture" are

$$\mathbf{A}_\phi = \frac{Q}{2\rho}\cos\theta(\rho), \mathbf{A}_\tau = \frac{1}{2}\cos\theta\partial_\tau\gamma(\tau);$$
$$\mathbf{H}_\tau = \Theta(\tau)\frac{2Q\tau^2}{(\rho^2+\tau^2)^2},$$
$$\mathbf{H}_\phi = \Theta(\tau)\frac{-2Q\rho\tau^2}{(\rho^2+\tau^2)^2}\partial_\tau\gamma(\tau),$$
$$\mathbf{H}_\rho = \Theta(\tau)\frac{2Q\rho\tau}{(\rho^2+\tau^2)^2} - \delta(\tau)\frac{Q}{2\rho}\frac{\rho^2-\tau^2}{\rho^2+\tau^2}. \quad (31)$$

One can easily confirm that

$$\nabla\cdot\mathbf{H} = Q2\pi\delta(\tau)\delta(\mathbf{r}), \quad (32)$$

representing a monopole of charge $Q$. So a space-time monopole event terminates a skyrmion. As $\rho,\tau$ approach infinity, $\mathbf{H}(\rho,\tau)$ becomes vanishly small. The action of this Euclidean space monopole charge is finite($a\sim 1$),

$$\mathcal{S}_{mon.} = \frac{1}{2g}\int d\tau d^2\mathbf{r}\mathbf{H}^2(\rho,\tau) = \frac{aQ^2}{16\pi g}; \quad (33)$$

($a$ is a constant of unity.) in addition, it has a phase factor following the Hopf term, which characterizes a rotation of the skyrmion during its duration

$$\gamma_B(\tau_0) = \frac{Q\mathcal{N}}{4}[\gamma(L_\tau) - \gamma(\tau_0)]. \quad (34)$$

Note that the term inside the bracket is not subject to the periodic condition imposed for the values of $\gamma$ in Eq.30. $\gamma_B$ obviously depends on the temporal coordinate at which the skyrmion is terminated, introduced explicitly as $\tau_0$ in Eq.34.

Taking into account this feature, the action of space-time monopoles $\{Q_\alpha, N_\alpha\}$ centered at $\mathbf{r}_\alpha, \tau_\alpha$ can be written as in Eq.C1. The most important feature thus is the space-time monopoles positioned at different points carry different phases $\gamma_B(Q_\alpha, \tau_\alpha)$. It is interesting that for Heisenburg antiferromagnetic spins at square lattices, the space-time monopoles carry spatially dependent Berry's phases which lead to spin -Peierls ground states [62].

A simple calculation in a saddle point approximation shows the energy density

$$\frac{E}{L^2} = -\delta(\mathcal{N})\frac{c\Delta_0}{\xi_0^2}\exp\left(-\frac{a}{16\pi g}\right)\cos(\chi_s). \quad (35)$$

Eq.35 indicates that only when $\mathcal{N}=0$, monopole events are liberated and the ground state is $\chi_s = n2\pi$, $n = 0,1,2....$ However, for $\mathcal{N}\neq 0$ the ground state is of arbitrary $\chi_s$ with infinite-fold degeneracy. The suppression of space-time monopoles is a unique feature of $QSDpSS$ which we are going to explore in some details.

To appreciate the robustness of Eq.35, we offer an alternative view from the perspective of the spectral flow. The absence of space-time monopoles in $QSDpSS$s is a direct consequency of a fermionic zero mode. Support for this conclusion comes from the following argument similar to that of the Witten's global anomaly [13]. The Hopf term in the action,

$$\mathcal{L}_{hopf} = \frac{\mathcal{N}}{4\pi^2}\int d^3x\epsilon_{\mu\nu\lambda}\big(\mathbf{A}_\mu + \frac{1}{2}\partial_t\beta\delta_{\mu,0}\big)\partial_\nu\mathbf{A}_\lambda, \quad (36)$$

suggests that the determinant of the fermionic operator

$$\mathcal{L} = \partial_\tau - \tau\cdot\mathbf{W}_0 - \mathcal{H}(\mathbf{W}_i), \quad (37)$$

transforms nontrivially under gauge transformations. Let us classify a gauge transformation as

$$\mathbf{W}_\mu(\mathbf{r},\tau) \to \mathbf{W}_\mu(\mathbf{r},\tau) + U_n^+\partial_\mu U_n, \quad (38)$$

where $U_n$ belongs to a nontrivial element of $\pi_3(S^2)$ group, or a Hopf texture. This has to be supplied with

$$\beta(L_\tau) - \beta(\tau+L_\tau) = n2\pi. \quad (39)$$

Under the transformation defined in Eq.38, the determinant acquires a negative factor due to the Hopf term

$$det\mathcal{L} \to (-1)^{\mathcal{N}n}det\mathcal{L}. \quad (40)$$

To see how this can happen, we analyze the spectral flow during different stages of gauge transformations. We introduce an additional parameter $\alpha$ to monitor the spectrum of the Lagrangian operator such that

$$\mathbf{W}_\mu(\mathbf{r},\tau;\alpha) = \begin{pmatrix} 0, & \alpha=0;\\ \mathbf{W}_\mu^n(\mathbf{r},\tau), & \alpha=1. \end{pmatrix} \quad (41)$$

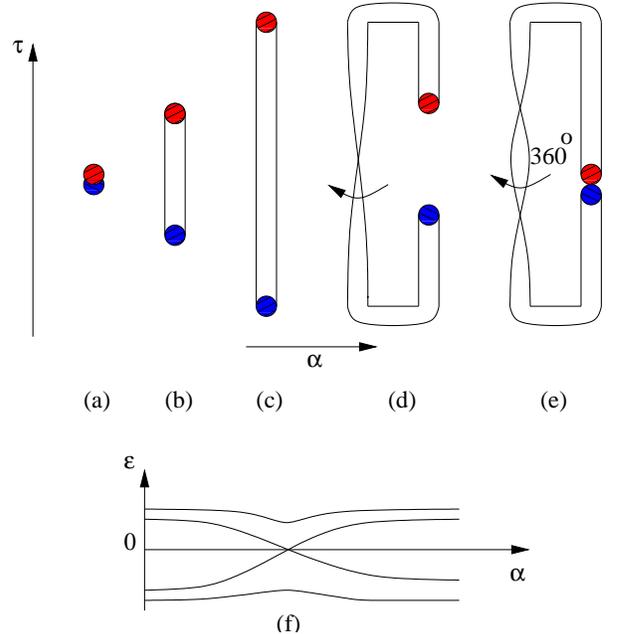

Fig.4 The different stages of gauge transformations in Eq.D1. a) $\alpha = 0$, b)$\frac{1}{2} > \alpha > 0$, c) $\alpha = \frac{1}{2}$, d)$1 > \alpha > \frac{1}{2}$, and e) $\alpha = 1$. A skyrmion (indicated by the tubes) is terminated at a monopole-antimonopole pair(indicated as filled circles). In f), a level crossing in the quasiparticle spectrum takes place at $\alpha = 1/2$.

At $\alpha = 0, 1$, the transformation is a pure gauge and the spectrum is given as

$$Det\mathcal{L}_0 = \prod_{n,\mathbf{k}} \left(\omega_n^2 + E(\mathbf{k})^2\right); \quad (42)$$

$\omega_n L_\tau = \pi(n+1/2)$. Recall $E(\mathbf{k}) = \sqrt{\Delta^2 + \epsilon(k)^2}$ and the spectrum of the Lagrangian operator is fully gapped.

In between, we have situations shown in Fig.3. We start with a skyrmion which is emitted from a monopole and terminated at an anti-monopole. Slowly we send the two monopoles to $\tau = \pm L_\tau/2$. At $\alpha = 1/2$, we will have two monopoles inserted into the boundaries of $2 + 1$ D. The gauge field $\mathbf{W}$ at $\alpha = 1/2$ precisely corresponds to that of two space-time monopoles. As two monopoles approach each other, the skyrmions are rotated by $360^0$ degrees.

To obtain a minus sign as indicated by the Hopf term, the spectrum has to cross each other once when the procedure specified in Eq.38 is carried out. Since the gauge fields are symmetric with respect to $\alpha = 1/2$, the spectrum is symmetric as well. The crossing has to take place at $\alpha = 1/2$ to give us a designated $-1$ contribution as suggested by the Hopf term.

This is the point where the gauge fields are most singular, defined by the monopole fields and the spectrum of the Lagrangian operator can be closed. The probability to have space-time monopoles is proportional to the ratio between the fermionic determinant in the presence of the monopole $det\mathcal{L}_1$ and that in a trivial background $det\mathcal{L}_0$,

$$\frac{det\mathcal{L}_1}{det\mathcal{L}_0} = \frac{\prod_n \Omega_n}{\prod_n \left(\omega_n^2 + E^2(\mathbf{k})\right)^2}. \quad (43)$$

Since the eigen value of the operator $\mathcal{L}_0$ is fully gapped, the zero energy mode $\Omega_0 = 0$ in the presence of a space-time monopole completely shuts down the monopole proliferation. This completes our argument. The argument only depends on the general principle of gauge invariance and not on the details of the gauge transformation.

It is important to mention that the action Eq.26 is valid when the time scale in the problem is shorter than $\Delta_s^{-1}$. The duration of the monopole should be understood as $\Delta_s$. The action for an individual space-time monopole in Eqs.33,34 is well-defined only when the perimeter $L_\tau$ is longer than $\Delta_s^{-1}$; otherwise, the thermal nucleation dominates. The crossover happens when

$$T \sim T^* = \frac{1}{\Delta_s(T)}. \quad (44)$$

It is therefore clear that in the *SDpSS* driven by quantum fluctuations, *QSDpSS*, the suppression of monopoles is extended into a finite temperature. On the other hand, for the *SDpSS* driven by thermal fluctuations, *TSDpSS*, $T$ is always higher than $T^*$. The Skyrmions are thermally nucleated; the topological phases are irrelevant for the discussions and the suppression is absent. The gauge fields and excitations in *QSDpSS*s are qualitatively different from the those in *TSDpSS*s.

### B. Zero energy skyrmions and quantum orders

To understand the issue of the conservation of skyrmion charges in *SDpSS*s, we first exploit at least three important intraconnected consequences of the zero modes. These three important features are vital for the discussions of quasiparticles in a SDpSS. First, the fluctuations of space-time monopole events per unit volumn defined as $<Q^2> = L^{-2} \partial^2 E(\chi_s)/\partial \chi_s^2$ are

$$<Q^2> = \delta(\mathcal{N}) \frac{c\Delta_0}{\xi_0^2} \exp(-\frac{a}{16\pi g}). \quad (45)$$

in *QSDpSS*s. Eq.45 shows that at any finite $\mathcal{N}$, all monopole events are suppressed due to destructive interferences between monopoles with different rotation angles $\gamma_B$. In *TSDpSS*s for the reasons mentioned at the end of the last subsection, the suppression is absent.

Second, Eq.45 indicates the conservation of the Skyrmion winding number at $\mathcal{N} \neq 0$ in a *QSDpSS*, that is in the absence of a spin rigidity. To see this, one first notices that the equation of motion for $\mathbf{H}_k$ is

$$\frac{\partial \mathbf{H}_k}{\partial t} = \epsilon_{ijk} \frac{\rho_s}{2m} \mathbf{n} \cdot \partial_j \mathbf{n} \mathbf{l} \cdot \partial_i \mathbf{n}. \quad (46)$$

Eq.46 demonstrates that the topological charge is conserved if singular space-time events are not allowed so that the product $\mathbf{n} \cdot \partial \mathbf{n}$ in the right hand side of the equation vanishes identically. And the topological charges fluctuate only when singular space-time events are permitted in the ground state.

This can be more explicit in $2D$. We define

$$C_{U(1)}(\{\mathbf{n}(\mathbf{r})\}) = \frac{1}{4\pi} \int dx dy \mathbf{H}_z, \quad (47)$$

as the total number of Skyrmions living on the 2D sheet. As shown before, a space-time monopole essentially connects a trivial vacuum to a Skyrmion configuration. Indeed, following Eqs.47,28,

$$\frac{\partial C_{U(1)}(\tau)}{\partial \tau} = \sum Q_m \delta(\tau - \tau_m), \quad (48)$$

where the surface contribution has been neglected since we are interested in the leading contribution to $C_{U(1)}$ per unit square. Following Eq.48, each monopole event causes a change in the topological charge $C_{U(1)}$ by one unit. The skyrmion charge is conserved in the absence of the space-time monopoles in $(2+1)$D.

Combining the results in Eqs.45, 48, we conclude that the skyrmion charge is indeed conserved in *QSDpSS*s. A skyrmion is a well-defined topological configuration even in the absence of spin stiffness, a remarkable result which only holds at $\mathcal{N} \neq 0$. Eq.43 indicates that the zero mode in instantons serves as a quantum protectorate to preserve the topological charge conservation. The energy of skyrmions originates from their interactions with spin fluctuations. For a size $\lambda$ skyrmion, following Eq.26 we arrive at an estimate

$$E_{sk.} = \frac{C\xi_2^2}{\lambda^2}\Delta_0(T) \qquad (49)$$

which vanishes as $\lambda$ goes to infinity.

Thirdly, this interesting destructive interference between different monopole events leads to very distinct behaviors of fields $\mathbf{F}_{\mu\nu}$. Formally, one can introduce an order parameter of the Wilson loop type to characterize the induced $U(1)$ fields. The Wilson-loop integral defined as

$$\mathcal{W}_{U(1)} = \langle \mathcal{P} \exp\left(i \oint \mathbf{A} \cdot d\mathbf{r}\right) \rangle \qquad (50)$$

has different asymtotical behaviors in the large loop limit in the presence or absence of the topological order in $C_{U(1)}$. When the topological charge $C_{U(1)}$ is conserved, the monopole-like instantons which connect topological different configurations are forbidden; the exponent in the Wilson loop integral is a linear function of the perimeter $L$ of the loop, i.e.

$$\mathcal{W}_{U(1)} = \exp(-LC_1). \qquad (51)$$

and the gauge fields are deconfining. Eqs.45,48 shows that in *QSDpSS*s, the gauge field is deconfining.

In *TSDpSS*s, on the other hand, $C_{U(1)}$ is not conserved; the instantons are allowed and one can confirm the exponent of the Wilson loop integral is a linear function of the area $S$ enclosed by the loop, i.e.

$$\mathcal{W}_{U(1)} = \exp(-SC_2). \qquad (52)$$

and the gauge fields are confining. For most of layered p-wave superconductors, given the gap energy at zero temperatures $\Delta_0 \approx 1K$ and the Fermi energy of order $1eV$, we conclude the superconductors are spin ordered at the zero temperature but could be in *TSDpSS*s (in the absence of spin-orbit coupling) as shown in Eq.24. *TSDpSS*s can be studied in many layered structure superconductors. We will be concerned with both situations, the *TSDpSS*s and *QSDpSS*s in the following discussion.

## IV. QUASI-PARTICLES IN A *SDPSS*

### A. $U(1)$ charges of Quasiparticles

We now employ the generalized Bogolubov-De Gennes equation to study the properties of quasiparticles in a *SDpSS*. In the presence of a topological configuration of $\mathbf{n}(\mathbf{r})$, it is convenient to introduce a gauge transformation

$$\Psi \to \exp(i\frac{\alpha(t)}{2}\sigma_3)\exp(i\frac{\beta(t)}{2}\tau_3)U_s(\mathbf{n})U_c(\chi)\Psi \qquad (53)$$

and work in a rotated representation; then one obtains a new Hamiltonian

$$H = \sigma_3\epsilon(i\nabla - \mathbf{A}_c - \mathbf{A}_s) + v_\Delta \sum_{i=1,2}\{\sigma_i\tau_3, i\partial_i - \mathbf{A}_{ci} - \mathbf{A}_{si}\}_+. \qquad (54)$$

Here $v_\Delta = \Delta_0/k_F$. The phase factors of $\alpha, \beta$ are to ensure antiperiodic temporal boundary conditions for quasi-particles, $\Psi(\tau) = -\Psi(\tau + L_\tau)$. $L_\tau$ is the perimeter along the direction of $\tau$; $\alpha(\tau) - \alpha(\tau + L_\tau) = n2\pi$, $\beta(\tau) - \beta(\tau + L_\tau) = n2\pi$.

Explicitly, we can have

$$\tilde{U}_s = u_s \begin{pmatrix} \cos\frac{\theta}{2}\exp\left(-i\frac{\phi}{2}\right), & -\sin\frac{\theta}{2}\exp\left(-i\frac{\phi}{2}\right) \\ \sin\frac{\theta}{2}\exp\left(i\frac{\phi}{2}\right), & \cos\frac{\theta}{2}\exp\left(i\frac{\phi}{2}\right) \end{pmatrix}. \qquad (55)$$

A direct calculation shows that

$$\mathbf{W}_\mu \cdot \sigma = i\tilde{U}_s^+(\theta,\phi)\partial_\mu\tilde{U}_s(\theta,\phi)$$
$$\mathbf{W}_\mu^1 = -\frac{1}{2}\sin\theta\partial_\mu\phi, \mathbf{W}_\mu^2 = -\partial_\mu\theta,$$
$$\mathbf{W}_\mu^3 = \frac{1}{2}\cos\theta\partial_\mu\phi + \frac{1}{2}\partial_\mu\alpha. \qquad (56)$$

It is clear that the $\mathbf{W}^3$ field represents the vector potential of $\mathbf{F}_{\mu\nu}$, the Pontryagin field introduced before,

$$\mathbf{W}_\mu^3 = \mathbf{A}_\mu + \frac{1}{2}\partial_\tau\beta(\tau)\delta_{\mu,0}. \qquad (57)$$

At last, the corresponding Lagrangian density is

$$\mathcal{L}_{BdeG} = \Psi^+\big(\partial_\tau - \sigma_3 A_0^{em} - \tau \cdot \mathbf{A}_{s0} - \mathcal{H}(\mathbf{A}_\mu^{em}, \mathbf{A}_{s\mu})\big)\Psi. \qquad (58)$$

Therefore, the interaction between a quasiparicle and spin fluctuations can be best characterized in terms of gauge fields introduced above. Following Eq.54, besides a usual electric magnetic charge defined with respect to $\mathbf{A}_{e.m.}$ fields, a *BdeG* quasiparticle also carries a unit $U(1)$-charge with respect to $\mathbf{A}_\mu$ fields and is minimally coupled with $\mathbf{F}_{\mu\nu}$. So both phase fluctuations and spin fluctuations affect the quasiparticles.

However, only the gradient of the phase, or superfluid velocity $\mathbf{v}_s$ enters Eq.54. Though phase fluctuations are enormous in 2D, the fluctuations in $\mathbf{v}_s$

$$<\mathbf{v}_s \mathbf{v}_s> = V_c^2 \frac{T}{\epsilon_F} \frac{\rho_0}{\rho_s(T)}, \qquad (59)$$

are small, except within the narrow Ginsburge region of the superconductor-metal transitions. We therefore conclude the BCS spectrum is not affected by phase fluctuations at temperatures far below $T_c$.

The spin fluctuations in $SDpSS$s on the other hand define all important properties of quasiparticles. A spin-$\frac{1}{2}$ excitation interacts with the spin fluctuations via gauge fields

$$H_I = \frac{1}{2} \int d\mathbf{r} \mathbf{J}_\mu \mathbf{A}_\mu$$
$$\mathbf{J}_\mu = e\delta(\mathbf{r} - \mathbf{r}^\alpha) \frac{\partial \mathbf{r}^\alpha}{\partial \tau}, J_0 = e\delta(\mathbf{r} - \mathbf{r}^\alpha). \qquad (60)$$

The energy of a spin-$\frac{1}{2}$ excitation therefore is

$$E = -\frac{1}{L_\tau} \ln \langle T \exp\big(-\frac{1}{2} \int_0^{L_\tau} H_I(\tilde{\tau}) d\tilde{\tau}\big) \rangle \qquad (61)$$

In terms of the induced connection fields,

$$E = -\frac{1}{L_\tau} \ln \langle \exp\big(-\frac{1}{2} \int_0^{L_\tau} d\mathbf{x}_0 \int_0^{L_2} d\mathbf{x}_2 \mathbf{F}_{02}\big) \rangle_{\{\mathbf{F}_{\mu\nu}\}}, \qquad (62)$$

and the average is taken over the partition function $Z(\{\mathbf{F}_{\mu\nu}\})$. Only the y-component of the electric field defined by $\mathbf{F}_{02}$ contributes to the energy. Say it differently, the energy of a $BdeG$ particle is determined by the Wilson loop integral of $\mathbf{A}$ fields. The evaluation of Eq.62 depends on the hidden topological long range order considered in recent preprints [53] (see also [28] for a general description on quantum orders). and provides all sorts of information on excitations. We will use this quantity to classify distinct structures of quasiparticles in $TSDpSS$s and $QSDpSS$s.

### B. The statistics of quasiparticles

In $QSDpSS$s, all monopole configurations are suppressed because of the destructive interference caused by Hopf terms and $<Q^2>=0$. The Wilson loop integral decays exponentially as a function of the perimeter of the loop. The interactions between $BdeG$ quasiparticles mediated by the topological fields $\mathbf{F}_{\mu\nu}$ are perturbative and the $BdeG$ quasiparticle energy is finite. But most importantly, in this case, the skyrmion winding number $\mathcal{C}_{U(1)}$ is a conserved quantity; skyrmions themselves carry $U(1)$ charges with respect to the fields of $\mathbf{A}_\mu$. This is indicated in Eq.18 if we introduce the skyrmion density-current density as

$$4\pi \mathbf{j}_\mu = \mathcal{N} \epsilon_{\mu\nu\eta} \partial_\nu \mathbf{A}_\eta \qquad (63)$$

and express the topological term in a form of minimal coupling $\mathbf{j}_\mu \mathbf{A}_\mu$. By minimizing the action of $\mathcal{L} + \mathcal{L}_{BdeG}$ with respect to $\mathbf{A}_0$, $\mathbf{A}^{em}$ and taking $\mathcal{N} = 1$ at a low temperature limit, we do obtain a saddle point equation $4\pi <\Psi^+ \tau_3 \Psi> = \mathbf{e}_z \nabla \times \mathbf{A}$, $<\Psi^+ \sigma_3 \Psi> = 0$. This shows that a skyrmion configuration carries a half spin but no charge. In other words, a spin $\frac{1}{2}$ but chargeless $BdeG$ quasiparticle is hosted by, or confined with a skyrmion, with the confinement mediated by the spin fluctuations. As a result, quasiparticles which are charge neutral with respect to $U(1)$ fields can be created. It is the vanishing energy of skyrmions in spin disordered superconducting liquids which makes these topological excitations relevant to the discussions of the low energy quasiparticles. In spin ordered liquids, the skyrmions' energy can be as high as the fermion energy and skyrmions are irrelevant for low energy quasiparticles.

For the study of a $BdeG$ particle confined within a skyrmion in $QSDpSS$s, consider a skyrmion with $n_W = 1$. In polar coordinates $(\rho, \phi)$, the director has a spatial distribution as

$$\mathbf{n}(\rho, \phi) = (\sin \theta(\rho) \sin \phi, \sin \theta(\rho) \cos \phi, \cos \theta(\rho)) \qquad (64)$$

and $\theta(\rho)$ is an arbitrary function of $\rho$, the asymptotics of which is $\theta(\rho = 0) = 0$ and $\theta(\rho \to \infty) = \pi$. The corresponding $\mathbf{A}$ field can be chosen as

$$\mathbf{A} = \frac{1 - \cos \theta(\rho)}{2\rho \sin \theta} \mathbf{e}_\phi, \nabla \times \mathbf{A} = \frac{\sin \theta(\rho)}{2\rho} \frac{\partial \theta(\rho)}{\partial \rho} \mathbf{e}_z. \qquad (65)$$

To facilitate the calculation of the fermionic spectrum hosted in a skyrmion, we further assume that $\theta(\rho) = \pi \Theta(\rho - \rho_0)$ is a step function of $\rho$. The $SU(2)$ field $\mathbf{W}^\alpha$ thus takes a simple form;

$$\mathbf{W}_i^3 = \mathbf{A}_i (i=1,2), \mathbf{W}_0^3 = 0;$$
$$\mathbf{W}_\mu^1 = \mathbf{W}_\mu^2 = 0. \qquad (66)$$

The gauge field $\mathbf{A}_\mu$ induced in the spin rotation above is simply the connection field of the Berry's phase. That is, $\int_\mathcal{C} d\mathbf{r} \mathbf{A} = \Omega_\mathcal{C}/2$ where $\Omega_\mathcal{C}$ is the solid angle spanned by a closed loop $\mathcal{C}$ swept by the director $\mathbf{n}$ on the unit sphere, when the particle moves around a skyrmion. In fact, Eq.65 represents a monopole field with the dirac string pointing at $-\mathbf{e}_z$ direction.

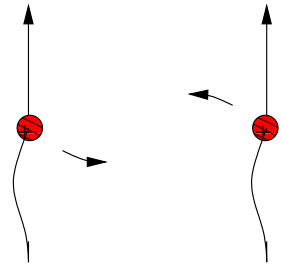

Fig.5 The composite structure of quasiparticles in *QS-DpSS*s. A bare *BdeG* quasiparticle (upward lines) and a skyrmion (doward lines) see each other as a dirac string pointing at $\mp \mathbf{e}_z$, following Eq.65.

When a spin-1/2 *BdeG* particle with spin up in the basis $U_s(\mathbf{n})(1,0)^T$ moves in a closed circle in a skyrmion defect, the acquired Berry's phase of $\pi[1-\cos\theta(\rho)]$ which approaches $2\pi$ at an infinity $\rho$. Similarly, a skyrmion moving around a spin-1/2 particle acquires a Berry's phase of $\pi\cos\theta(\rho)$. The total phase involved in moving the composite around the other one is $2\pi$. Consequently, under the interchange of coordinates, the two-body wavefunction of composite quasiparticles acquires an additional $\pi$ phase because of hosting skyrmions,

$$\Psi(\mathbf{r}_1, \mathbf{r}_2) = \Psi(\mathbf{r}_2, \mathbf{r}_1). \tag{67}$$

Eq.67 also follows the linking number theorem for skyrmions [17]. These composite quasi-particles are therefore Bosons. We also observe at $\epsilon = 0$, *BdeG* quasi-particles are charge neutral with respect to an *em* field; they also carry zero $U(1)$ charges so to minimize the interaction between composite excitations.

The life time of the composited quasiparticles is determined by the minima of the life time of skyrmions which is limited by the quantum nucleation rate, and the usual inelastic scattering time. It is important that because of the destructive interferences between rotated skyrmions, the rate is suppressed to zero in a *QSDpSS* even at a finite temperature and the skyrmion charge $C_{U(1)}$ is conserved as emphasised in a few occasions in the paper.

In *TSDpSS*s, the Hopf term is irrelevant because $L_\tau \ll \Delta_s$; following Eq.C1, the partition function of a monopole configuration $\{\mathbf{x}^m\}$ is

$$Z = L_\tau \sum_N \frac{1}{N!} \sum_{Q_\alpha = \pm 1} \int \prod_{\alpha=1}^N d\mathbf{r}_\alpha \exp\left(-S_Q^{2d} + i\chi_s Q_\alpha\right). \tag{68}$$

And $S_Q^{2d} \sim \Delta/T$ represents a thermal activation factor.

Eq.68 suggests that in 2d *TSDpSS*s, the skyrmions are condensed and $<Q^2> \neq 0$. The topological charge $C_{U(1)}$ is not conserved. As emphasised before, when the particle moves around a condensed skyrmion of an arbitrary size, it picks up a Berry's phase between zero and $2\pi$. The probability for a particle to propagate at a large distance $L$ is

$$G(1,2) \sim \sum_m \exp(-i\Gamma_m), \Gamma = \sum_i \int_m d\mathbf{r} \cdot \mathbf{A}(\mathbf{r}, i); \tag{69}$$

$A(i)$ is the vector potential due to the ith skyrmion and $\Gamma_m$ is the total phase acquired by the quasiparticle moving along path $m$. Given that skyrmions are randomly nucleated in the 2D space and assuming the paths are cigar-like ones, we estimate

$$-\ln G(1,1) \propto L_m, \tag{70}$$

with $L_m$ being the length of a characteristic loop $m$. The estimate shows the particle is localized because of the destructive interferences between different paths in the presence of thermally nucleated skyrmions. Or say it differently, a bare *BdeG* is infinitily massive because of the skyrmion cloud.

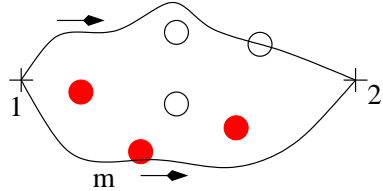

Fig.6 Quasiparticles in a cloud of skyrmions. The filled circles stand for the positive skyrmions and empty ones for negative skyrmions. The Berry's phase for a quasi-particle moving along a skyrmion could be between zero and $2\pi$.

For a similar reason, the expectation value in Eq.62, which is equal to the Wilson loop integral of a gauge field, decays exponentially with an exponent proportional to $PL_\tau L_2$; $P = \exp(-S_Q^{2d})$ is the thermal activation rate. The energy of a *BdeG* quasiparticle is proportional to the size of the system. And the bare *BdeG* particles could have been confined in a *TSDpSS*. In general, in a *TS-DpSS*, the mechanism of having zero energy skyrmions is absent and the $C_{U(1)}$ is not conserved. Bare *BdeG* quasiparticles interact with a linear potential and form bound states.

**Tab. 1 Comparison of quasiparticles**

|         | Compos.        | Spin | $Q_{e.m.}$ | $Q_{U(1)}$ | DoS    | Sta.  |
|---------|----------------|------|-----------|------------|--------|-------|
| *SOpSS* | bare *BdeG*    | 1/2  | 0*        | 1          | gapped | Fermi |
| *QSDpSS*| *BdeG*+ Skyr.  | 1/2  | 0         | 0          | gapped | Bose  |
| *TSDpSS*| *BdeG*+ *BdeG* | 0,1  | 0         | 0          | gapped | Bose  |

*This is for the charge at $\epsilon(\mathbf{k}) = 0$. $Q_{e.m.}$, $Q_{U(1)}$ are charges defined with respect to fields $\mathbf{A}_{e.m.}$ and $\mathbf{A}$.

### C. Collective spin excitations

As emphasised in the previous section, in *TSDpSS*s, the $U(1)$ fields are confining. An excitation carrying a unit charge is confined and is forbidden. The collective excitations have to be "charge neutral" with respect to **A**. A spin-1/2 $\eta$ which carries a unit charge is absent in the excitation spectrum; only pairs of $\eta$ which carry zero $U(1)$ charges exist as physical excitations in a form of the usual spin wave excitations. This is what happens in $TSDpSS$s.

However, the topological term changes the nature of interactions mediated by **A** fields in $QSDpSS$s. In fact, an $\eta$ quantum is bound with a skyrmion such that the bound state becomes a fermion. Each bosonic spin wave excitation is fractionalized into two elementary fermionic spinors hosted in skyrmions in the spin disordered limit. Each spinor-skyrmion composite carries a half spin but no $U(1)$ charge, by contrast to a bare $\eta$ excitation.

**Tab. 2 Comparison of collective spin excitations**

|  | Compos. | Spin | $Q_{e.m.}$ | $Q_{U(1)}$ | DoS | Sta. |
|---|---|---|---|---|---|---|
| $SOpSS$ | Spin waves | 1 | 0 | 0 | gapless | Bose |
| $QSDpSS$ | Spinor + Skyr. | 1/2 | 0 | 0 | gapped* | Fermi |
| $TSDpSS$ | Spinor + Spinor | 0, 1 | 0 | 0 | gapped* | Bose |

* However, it is unclear how the magnitude of the gap and the range of spin correlation are renormalized by the topological term.

## V. TOPOLOGICAL EXCITATIONS

For the sake of completeness, I am also listing some properties of vortices. The linear defects of a symmetry broken state with an internal space $\mathcal{R} = [S^1 \times S^2]/Z_2$ have been recently discussed extensively in the context of Bose-Einstein condensates of $^{23}Na$ [48,53]. In a $SOpSS$, the linear defects are superpositions of $hc/4e$ vortices and $\pi$-disclinations because of the $Z_2$ symmetries in the problem. An individual $\pi$-disclination or $hc/4e$ vortex carries a branch cut along which the phase jumps by $\pi$. The branch cut emitted by a $hc/4e$ vortex can end only at a $\pi$-disclination, resulting in a linear potential between these objects and therefore confinement. And a bare $hc/4e$ vortex is forbidden because of the catastrophe of a cut. In a $SDpSS$, however, $hc/4e$ vortices can exist by its own right and are elementary excitations.

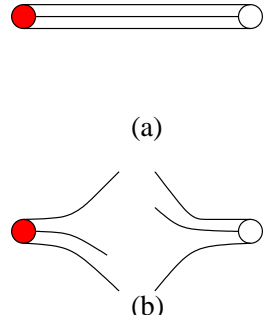

Fig.5 a) The confinement of $\pi$-disclinations(patterned circles) and half vortices(empty circles) in a $SOpSS$ by a string, or a branch cut; b) the liberation of half-vortices in a $SDpSS$.

The $SDpSS$ discussed here has the following order parameters:

$$< \hat{\Delta} > = 0, Tr < \hat{\Delta}\hat{\Delta} > \neq 0, < \exp(i\chi) > \neq 0. \quad (71)$$

The existence of $SC^*$ with $< \hat{\Delta} > = 0$, $Tr < \hat{\Delta}\hat{\Delta} > \neq 0$, $< \exp(i\chi) > = 0$, and other fractionalized states examined in [41] is beyond the model studied here. Physically, the $SDpSS$ has Josephson oscillations of $2eV$ frequency while in $SC^*$ the frequency is $4eV$.

In the presence of spin-orbital couplings, the mean field solution indicates that the director of **n** points along $\pm \mathbf{e}_z$ direction and the internal space is $[Z_2 \times S^1]/Z_2$. However, at the energy scale higher than the spin-orbit coupling ones, **n** would be free to rotate on a two-sphere. The spin order-disorder transition still takes place at a finite temperature below the superconductor-metal transition temperature $T_c$ when the spin-orbit scattering rate is much smaller than $\Delta_0(0)$. As the spin is disordered, skyrmions are again zero energy objects and the discussions on the quasiparticles are still valid.

## VI. CONCLUSION

In conclusion, we also would like to remark that some aspects of the quasiparticles in spin disordered superconductors considered here reminisce the chiral-bag defect model for the nucleon [64,65]. The presence of spin-1/2 bosonic chargeless $BdeG$ excitations in a $SDpSS$ is an example of fermi number fractionalization; it belongs to the same class phenomenon as the mid-gap quasiparticles hosted in domain wall excitations in one dimension polyacetylene [6]. and the statistical transmutations proposed in some magnetic models [40,18,23]. Though in this case all the excitations in the spin rotation invariant superconducting state remain gapped, the spectral property should be very different from that in conventional spin triplet superconductors when the $S^2$ symmetry is broken [66].

## VII. ACKNOWLEDGEMENT

It is my pleasure to thank P. W. Anderson, E. Demler, K. Schoutens, P. Wiegmann, F. Wilczek and X. G. Wen for useful discussions. Finally I am grateful to B.Y. L. Luo for encouragement and supports.

## APPENDIX A: DERIVATION OF EQ.13 USING SUB-LATTICES

Eq.13 can also be obtained by considering two sublattices $A$ and $B$ with $u(\mathbf{\Omega}_{i1})$ and $v(\mathbf{\Omega}_{i2})$ labeling the state at site $i$ of the lattice $A$ and $B$. At each site the total spin of $A$ and $B$ is one. One can use the results for $u(\mathbf{\Omega})$ in Eq.5 to obtain the effective field

$$F_{\mu\nu} = F_{\mu\nu}^A - F_{\mu\nu}^B$$
$$F_{\mu\nu}^{A,B} = \frac{1}{2}\epsilon_{\alpha\beta\gamma}\mathbf{\Omega}_{1,2}^\alpha \cdot \frac{\partial \mathbf{\Omega}_{1,2}^\beta}{\partial x_\mu}\frac{\partial \mathbf{\Omega}_{1,2}^\gamma}{\partial x_\nu}. \quad (A1)$$

Taking into account $\Omega_{1,2} = \mathbf{n} - \frac{1}{2}$, one obtains the results previously derived in Eq.13. From here it is clear how the contributions from $u(\mathbf{\Omega}_1)$ and $v(\mathbf{\Omega}_2)$ cancel with each other.

## APPENDIX B: DERIVATION OF COEFFICIENTS IN EQ

.23

$$\mathcal{T}_0 = \frac{\pi kT}{V}\sum_{\omega_n,\mathbf{k}}\frac{-\omega_n^2 + \epsilon^2(\mathbf{k})}{(\omega_n^2 + E(\mathbf{k})^2)^2},$$
$$\mathcal{S}_{ab} = \frac{\pi kT}{V}\sum_{\omega_n,\mathbf{k}}\frac{\Delta_0^2 \mathbf{v}_a \mathbf{v}_b}{(\omega_n^2 + E(\mathbf{k})^2)^2},$$
$$\mathcal{T}^{\alpha\beta} = \delta_{\alpha\beta}\frac{\pi kT}{V}\sum_{\omega_n,\mathbf{k}}\frac{\Delta_0^2}{(\omega_n^2 + E(\mathbf{k})^2)^2}. \quad (B1)$$

$\omega_n = (2n+1)\pi kT$, $n = 0, 1, 2...$ are the Masubara frequencies. Eq.B1 leads to Eq.23.

$F_{1,2}$ in Eq.23 can be calculated as

$$F_1(\mathbf{k},T) = -\frac{\partial}{\partial\epsilon^2(\mathbf{k})}\frac{2\Delta_0^2(T)}{E(\mathbf{k})}\tan(\frac{E(\mathbf{k})}{2kT})$$
$$F_2(\mathbf{k},T) = -\frac{\Delta^2}{E^3(\mathbf{k})} + \frac{\partial}{\partial\epsilon(\mathbf{k})}\left(\frac{\epsilon}{E(\mathbf{k})} + \frac{E(\mathbf{k})}{\epsilon}\right). \quad (B2)$$

## APPENDIX C: THE ACTION FOR SPACE-TIME MONOPOLES

The action for space-time monopoles is

$$Z = \sum_{N=0,}^{+\infty}\frac{1}{N!}\prod_{\alpha=1}^{N}c\int d\mathbf{r}_\alpha d\tau_\alpha \sum_{Q_\alpha=\pm 1}$$
$$\exp\left(i\sum_\alpha \gamma_B(Q_\alpha,\tau_\alpha) + i\sum_\alpha \chi_s Q_\alpha - S_Q\right),$$
$$S_Q = \sum_{\alpha\neq\beta}\frac{Q_\alpha Q_\beta}{2g(|\mathbf{r}_\beta-\mathbf{r}_\alpha|^2 + (\tau_\beta-\tau_\alpha)^2)^{1/2}} + \sum_\alpha \frac{Q_\alpha^2}{2g}. \quad (C1)$$

Note that the first phase factor in the exponent arises from the Hopf term, distinguishing this system from the usual compact $U(1)$ theory. $\chi_s$ is introduced as a Sine-Gordon field coupled to a monopole charge $Q_\alpha$.

## APPENDIX D: GAUGE FIELDS DURING THE GAUGE TRANSFORMATION IN EQ.38

The gauge fields in between ($0 < \alpha < 1$) can be choosen to be

$$\mathbf{H}_\tau = \frac{\tau-\tau_1}{2(r^2+(\tau-\tau_1)^2)^{3/2}} + \frac{\tau_2-\tau}{2(r^2+(\tau-\tau_2)^2)^{3/2}}$$
$$\mathbf{H}_x = [x\cos\beta(\tau) + y\sin\beta(\tau)]\times$$
$$[\frac{1}{2(r^2+(\tau-\tau_1)^2)^{3/2}} - \frac{1}{2(r^2+(\tau-\tau_2)^2)^{3/2}}]$$
$$\mathbf{H}_y = [-x\sin\beta(\tau) + y\cos\beta(\tau)]\times$$
$$[\frac{1}{2(r^2+(\tau_1-\tau)^2)^{3/2}} - \frac{1}{2(r^2+(\tau-\tau_2)^2)^{3/2}}]$$
(D1)

and $\beta(\tau_1) - \beta(\tau_2) = n2\pi$. $\tau_1(\alpha) < \tau_2(\alpha)$ varies from $0(\alpha=0)$ to $\mp L_\tau/2$ ($\alpha=1/2$) and back to 0 at $\alpha=1$. In this configuration, the skyrmion charge varies as a function of $\tau$,

$$C_{U(1)} = \frac{1}{2}[\Theta(\tau-\tau_1) - \Theta(\tau_1-\tau)$$
$$+\Theta(\tau_2-\tau) - \Theta(\tau-\tau_2)]. \quad (D2)$$